\documentclass[final,1p,times]{elsarticle} 
\usepackage{graphicx} 
\usepackage{amssymb} 
\usepackage{amsthm} 
\usepackage{lineno} 
\usepackage{amsmath,amssymb,bm}

\journal{Nuclear Physics A} 
\begin{document} 

\def\Nfour{\mathcal N\,{=}\,4}
\def\Ntwo{\mathcal N\,{=}\,2}
\def\Nc{N_{\rm c}}
\def\Nf{N_{\rm f}}
\def\x{\bm x}
\def\q{\bm q}
\def\f{\bm f}
\def\v{\bm v}
\def\t{\bm t}
\def\S{\mathcal S}
\def\T{\mathcal T}
\def\O{\mathcal O}
\def\E{\mathcal E}
\def\p{\mathcal P}
\def\H{\mathcal H}
\def\K{\mathcal K}
\def\w{\omega}
\def\uh{u_h}
\def\del{\nabla}
\def\eps{\hat \epsilon}
\def\inf{\epsilon}
\def\cs{c_{\rm s}}
\def\tensor{ \overleftrightarrow }
\def\Deltax{\Delta x_{\rm max}}
\def\C{\mathcal C}
\def\u{{\mathcal U}}
\def\uo{{u_*}}
\def\r{{\xi}}
\def\rr{{\xi}}
\def\to{{t_*}}
\def\xendpt{x_{\rm endpt}}
\def\xendpt{\mathcal X}
\def\t{\tau}

\begin{frontmatter} 


\title{Gauge/gravity duality and jets in strongly coupled plasma}

\author{Paul~M.~Chesler}

\address{
Department of Physics,
University of Washington,
Seattle, WA 98195-1560, USA
}

\begin{abstract} 
We discuss jets in strongly coupled $\Nfour$ supersymmetric Yang-Mills plasma and their dual gravitational description.
\end{abstract} 

\end{frontmatter} 



\section{Introduction}

The discovery that the quark-gluon plasma produced
at RHIC behaves as a nearly ideal fluid \cite{Shuryak:2004cy}
has prompted much interest into the dynamics of strongly coupled plasmas.
Energetic partons produced in hard processes during 
the early stages of heavy ion collisions
can traverse the resulting fireball and deposit their energy and
momentum into the medium.  Analysis of particle correlations
in produced jets can provide useful information about the dynamics
of the plasma including the rates of energy loss and momentum broadening
\cite{Leitch:2006ex,CasalderreySolana:2006sq},
as well as the speed and attenuation length of sound waves \cite{CasalderreySolana:2006sq}.

Studying the dynamics of jets in QCD from first principles --- from the initial 
hard process which created the jet to energy loss, showering and thermalization ---  is a 
challenging task.   
Because of the lack of theoretical tools available for analyzing
strongly coupled dynamics in QCD, it is useful to have 
a toy model which is qualitatively similar to QCD but where strongly coupled dynamics can be treated in 
a controlled manner.  A class of such toy models are non-Abelian theories with gravitational duals 
\cite{Maldacena:1997re, Witten:1998qj}.
The most widely studied example is that of strongly coupled $\Nfour$
supersymmetric Yang-Mills theory (SYM).
The deconfined plasma phases of QCD and SYM share many qualitative properties.
For example, both theories describe non-Abelian plasmas
with Debye screening, finite spatial correlation lengths,
and long distance dynamics described by neutral fluid hydrodynamics.
When both theories are weakly coupled, appropriate comparisons
of a variety of observables show rather good agreement
\cite{CaronHuot:2006te,Huot:2006ys,Chesler:2006gr}.
This success, combined with the lack of alternative techniques
for studying dynamical properties of QCD at temperatures where
the plasma is strongly coupled, has motivated much interest in using
strongly coupled SYM plasma as a model for QCD plasma
at temperatures $T$ of a few times $\Lambda_{\rm QCD}$
(or $1.5 \, T_{\rm c} \lesssim T \lesssim 4 \, T_{\rm c}$).

Despite their qualitative similarities at finite temperature, 
it should be emphasized that dynamics in SYM are not a controlled approximation 
to those in QCD.  Moreover, in regimes where asymptotic freedom may be important the qualitative
dynamics in SYM may not resemble those in QCD.  This is certainly the case during the 
initial hard process which creates a jet.  In QCD, asymptotic freedom guarantees that for sufficiently 
high energy processes, the early time dynamics will be weakly coupled.  Therefore, in QCD the 
production of high energy jets will be governed by perturbative physics.
This should be contrasted with SYM, where the coupling doesn't run and early time dynamics are
just as strongly coupled as late time dynamics.

Because of the above point, we chose to focus of quantities which are largely 
insensitive to the initial event which created the jet.  
Our motivation is similar to that of Ref.~\cite{Liu:2006ug}, in which
weak coupling physics in asymptotically free QCD is envisioned
as producing a high energy excitation, whose propagation through
the plasma is then modeled by studying the behavior of a similar
type of excitation in a strongly coupled SYM plasma.
As SYM contains only adjoint matter, fundamental fields must be added to the theory
in order to study the propagation of fundamental representation excitations.
Within the context of gauge/gravity duality, a natural choice is to add a fundamental 
$\Ntwo$ hypermultiplet to SYM and study $\Ntwo$ excitations \cite{Karch:2003nh}.  

By studying $\Ntwo$ excitations in strongly coupled SYM plasma we can then answer
the following simple set of important questions. 
How far can an energetic jet in a plasma propagate before it thermalizes?  
Where along its trajectory does a jet prefer to deposit energy and momentum?
How well is the radiation created by the jet modeled by hydrodynamics? 
Answering this last question is important as hydrodynamics
is a much simpler theory to work with than any complete microscopic theory and
it applies equally well to QCD and SYM.

\section{Jets and their dual gravitational description}

Energetic quarks moving through a plasma are quasi-particles ---
they have a finite lifetime which can be long compared to the inverse
of their energy.
Because of this, some care is needed in defining observables such as the instantaneous energy 
loss rate or the penetration depth.  
At weak coupling, where a perturbative analysis is applicable, 
an energetic quark scattering off excitations in the medium can emit gluons
which may subsequently split into further gluons or quark-antiquark pairs.
An energetic quark may also annihilate with an antiquark in the medium.
A natural question to consider then is which quark should one follow when computing either the position of the 
total excitation or its instantaneous energy loss rate?
Once a quark has emitted a $q \bar q$ pair,
or annihilated with an antiquark,
it becomes ambiguous which quark was the original one.
This issue is most cleanly avoided if one focuses attention not on some
(ill-defined) ``bare quark'', but rather on the locally conserved charges of the
entire dressed excitation.

In both QCD and SYM coupled to a fundamental $\Ntwo$ hypermultiplet, 
there are two conserved quantities which are useful for characterizing a fundamental representation excitation.  
These are simply the baryon number (or quark number) current $J^\mu_{\rm baryon}$ and the total stress tensor 
for the system $T^{\mu \nu}$.
Even though $q \bar q$ pairs can be produced by an energetic quark
traversing the plasma, conservation of
$J^{\mu}_{\rm baryon}$ and $T^{\mu \nu}$ implies that the total
baryon number and four momentum of the state will remain constant.
For an energetic, well collimated jet, the baryon density can remain
highly localized for a long period of time. It is the collective excitation with localized baryon density
which we will refer to as a dressed quark, or for the sake of brevity, simply as a quark.

\begin{figure}[t]
\centering
\includegraphics[scale=0.26]{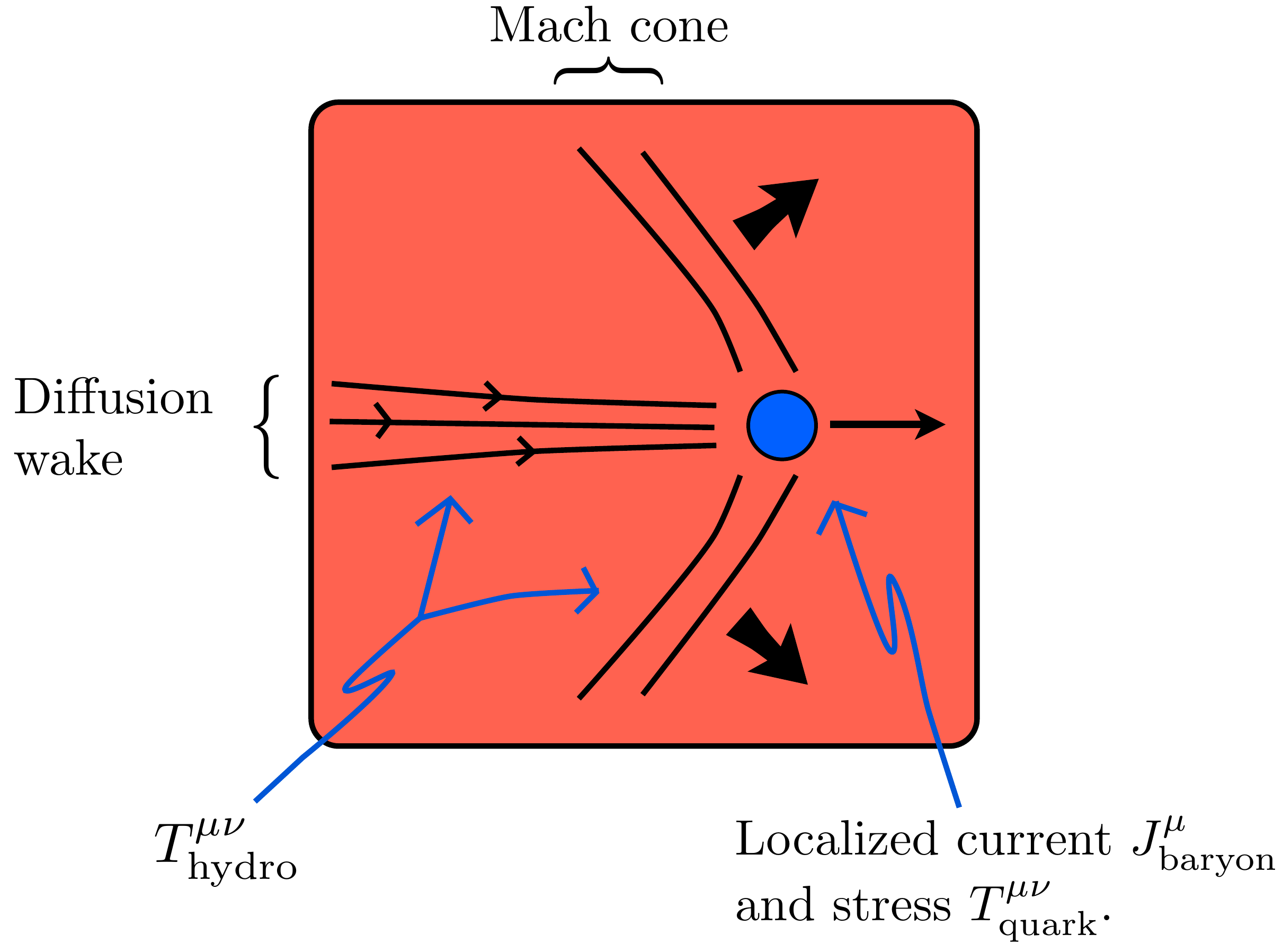}
\caption
    {\label{jetbigpicture}
    A cartoon depicting the large scale structure of an energetic quark moving through a plasma.
     }
\end{figure}

With the above points in mind, let us for the moment focus not on the microscopic details 
needed to fully describe a quark moving through a plasma but rather on the large scale structure.
Fig.~\ref{jetbigpicture} shows a cartoon depicting the large scale picture of a quark moving
though a plasma.  
The quark itself is a localized object whose location is specified by the localized baryon density.
Specifically, the trajectory of the quark is \textit{defined} to be
\begin{equation}
\label{centroid}
\bm x_{\rm quark}(t)  \equiv
\frac{1}{Q} {\int d^3 x \> \bm x \, \rho(t,\bm x)},
\end{equation}
where $\rho \equiv J^0_{\rm baryon}$ and $Q\equiv \int d^3 x \, \rho(t,\bm x)$ is the baryon number of the quark.
For very energetic quarks the stress tensor will also be highly
peaked in the vicinity of $\bm x_{\rm quark}(t)$.  
As the quark moves through the plasma, it will experience frictional drag forces and consequently, energy and momentum will
be transported from the vicinity of $\bm x_{\rm quark}(t)$ to the far zone.  
Because SYM is dissipative,
gradients in the stress tensor necessarily become smaller and smaller at larger distances,
and the far-zone dynamics of emitted radiation are governed by hydrodynamics. 
In the long wavelength limit the total stress tensor may be written
\begin{equation}
\label{stressdecomp}
T^{\mu \nu} \approx T^{\mu \nu}_{\rm hydro} + T^{\mu \nu}_{\rm quark},
\end{equation}
where $T^{\mu \nu}_{\rm quark}$ is localized about $\bm x_{\rm quark}(t)$
and $T^{\mu \nu}_{\rm hydro}$ satisfies 
the constituent relations of hydrodynamics together with the energy-momentum conservation equation
\begin{equation}
\label{hcons}
\partial_\mu T^{\mu \nu}_{\rm hydro}(\bm x,t) = J^{\nu}(\bm x,t), \ \ \ \ J^\nu(\bm x, t)  \equiv - \partial_\mu T^{\mu \nu}_{\rm quark}(\bm x, t).
\end{equation}

The \textit{effective source} $J^\mu$ for hydrodynamics is localized at the position of the quark and 
takes into account all microscopic dynamics in the near-zone relevant for the determination of the 
far-zone hydrodynamic behavior.  In general its determination will require a complete
microscopic calculation.  However, just by energy-momentum conservation, 
one crucial property the effective source must satisfy is that 
its integral over space must give the rate that the quark loses 
four-momentum:
\begin{equation}
\label{fdef}
 \int d^3 x \, J^\mu (\bm x,t)  = - \frac{d}{dt} \int d^3 x \, T^{0 \mu}_{\rm quark}(\bm x,t) \equiv f^\mu(t) ,
\end{equation}
where $f^\mu(t)$ is minus the drag force acting on the quark, and thus the rate that four-momentum
is transfered to hydrodynamic modes.%
\footnote
  {
  \label{fave}
  More precisely, as hydrodynamics is an effective theory valid on scales much larger
  than the mean free path, the integral of the effective source 
  must only give the microscopically time-averaged drag force acting on the quark.
  }
Operationally, the instantaneous drag force can be obtained by 
enclosing the quark's baryon density in a sphere and computing the total outward flux of four
momentum as a function of time.  

Just as in electromagnetism, where local sources can be expanded in terms of point-like multipole moments
(\textit{i.e.} derivatives of delta functions) with each moment producing a particular falloff of the fields with distance, 
the long distance hydrodynamic response of an energetic quark traversing a plasma
can be systematically expanded in inverse powers of distance. 
In doing so, the leading long distance behavior of the hydrodynamic response is determined
by the source's monopole moment \cite{Chesler:2007sv}.  More explicitly, at leading order the source
may be approximated
\begin{equation}
\label{leadingsource}
J^\mu(\bm x,t) \approx f^\mu(t) \, \delta^3(\bm x - \bm x_{\rm quark}(t) ).
\end{equation}
We note that the coefficient of the delta function is constrained via Eq.~(\ref{fdef}).
Together with the hydrodynamic constituent relations, Eqs.~(\ref{hcons}) and (\ref{leadingsource})
show that the leading order behavior of the 
far-zone stress tensor is determined solely by the quark's trajectory $\bm x_{\rm quark}$, $f^\mu$ and the plasma's viscosity.

In general there are two possible hydrodynamic modes an energetic quark can excite.  These modes are a 
sound mode and diffusion mode.  The diffusion mode trails behind the quark and slowly 
broadens with distance.  This mode, symbolized in Fig.~\ref{jetbigpicture} by the trailing flow
lines, consists of laminar flow in the same direction as the quark.  For supersonic motion
the sound mode manifests itself as a Mach cone or sonic boom.  This mode is 
symbolized in Fig.~\ref{jetbigpicture} by the outward moving conical wake emanating 
from the quark.

For heavy quarks whose motion
is not ultrarelativistic, $f^\mu(t)$ only depends on the temperature of the plasma, the 't Hooft coupling
and the quark's velocity \cite{Herzog:2006gh, CasalderreySolana:2006rq, Gubser:2006bz}.   
However for ultrarelativistic quarks $f^\mu(t)$ can be sensitive to the initial conditions which created the quark
\cite{Chesler:2008uy}.
As the weakly coupled hard process which creates a jet in QCD
is very different than that of strongly coupled SYM,
for the case of light quarks --- whose motion is always ultrarelativistic ---
it is desirable to have another observable which is related to energy loss but not
very sensitive to initial conditions.  One quantity to consider is the maximum distance $\Deltax(E)$
which a quark with initial energy $E$ can travel.  One can imagine fixing the quark's energy
and varying all other degrees of freedom characterizing the state and looking for the particular configuration
which maximizes the penetration depth.  Therefore, essentially by construction, the initial conditions
associated with quarks that maximize the penetration depth at fixed energy
are much more constrained.

To define the penetration depth, we imagine measuring
$\bm x_{\rm quark}(t)$ at some early time $\to$.
We then define the penetration depth $\Delta x$ as
$
\Delta x \equiv |\bm x_{\rm quark} (\infty) -\bm x_{\rm quark}(\to)| \,.
$
To make the quantity $\Delta x$ meaningful,
we also need to measure the
quark's energy at time $\to$.  If the quark has been moving for some time
prior to $\to$, it will have deposited energy into the plasma ---
we must disentangle the
energy deposited in the plasma from the remaining energy of the quark itself.
In the limit where the quark has an arbitrarily large energy which
is arbitrarily localized about $\bm x_{\rm quark}$, separating
the quark's energy from the energy transfered to the plasma
will be unambiguous.  We consider only this limit in this paper.

We now discuss jets in SYM from the $5d$ gravitational perspective.
The $10d$ geometry corresponding to an infinite equilibrium SYM plasma is
the product of a $5d$ sphere and the $5d$ AdS-Schwarzschild (AdS-BH)
geometry, whose metric is
$
    ds^2 = \frac{L^2}{u^2}
   [-f(u) \, dt^2 + d \x^2 + du^2/f(u) ],
$
where $f(u) \equiv 1-(u/u_h)^4$ and $L$ is the AdS curvature radius.
The AdS-BH geometry contains a $4d$ boundary
at radial coordinate $u = 0$. This boundary has the geometry of Minkowski space and is where
the $4d$ dual field theory should be thought of as residing.
The AdS-BH geometry also contains a translationally invariant event horizon at $u = u_h$ and has
a corresponding Hawking temperature
$T = 1/\pi u_h$.  

The addition of a $\Ntwo$ hypermultiplet to SYM
is accomplished in the gravitational setting by adding a D7 brane to the $10d$ geometry
\cite{Karch:2003nh}.
The D7 brane
fills a volume of the AdS-BH geometry which extends from the boundary
at $u = 0$ down to 
maximal radial coordinate $u_m$.  
The bare mass $M$ of the hypermultiplet is proportional to
$1/u_m$ \cite{Herzog:2006gh}.
Therefore, for massless quarks the
D7 brane fills all of the five-dimensional AdS-BH geometry and for very 
heavy quarks the D7 brane ends very close to the boundary at $u = 0$.
Open strings which end on the D7 brane correspond to \textit{dressed} quarks in the 
dual field theory.  

In general, to study the dynamics of $q \bar q$ pairs in the field theory
one must study dynamics of strings in the full $10d$ geometry.  The reason for this
is that in general string endpoints have non-trivial trajectories on the $S^5$.
However, in the limit of large quark mass and very small quark mass 
it is consistent to set the motion on the $S^5$ to vanish \cite{Chesler:2008uy}.
Because of this simplification and because any additional motion of the string on the $S^5$ will only
add to the energy of the string without affecting its stopping distance in the spatial direction, 
we choose to study only the dynamics of very heavy and massless quarks in this paper.  

The motion of strings corresponding to very heavy or very light quarks
is quite different.  For the case of very heavy quarks, the D7 brane 
ends at $u \rightarrow 0$ and string endpoints always remain 
close to the boundary.  For very light quarks, 
the D7 brane fills the entire AdS-BH geometry and string endpoints are allowed 
to fall unimpeded towards the black hole \cite{Chesler:2008uy}.
However, in either case the late-time spatial motion of the string
endpoint always ceases and in either case the total distance 
traveled by the endpoint can be made arbitrarily far by giving the string
more momentum in the spatial directions \cite{Chesler:2008uy}.

\begin{figure}[h!]
\centering
\includegraphics[scale=0.35]{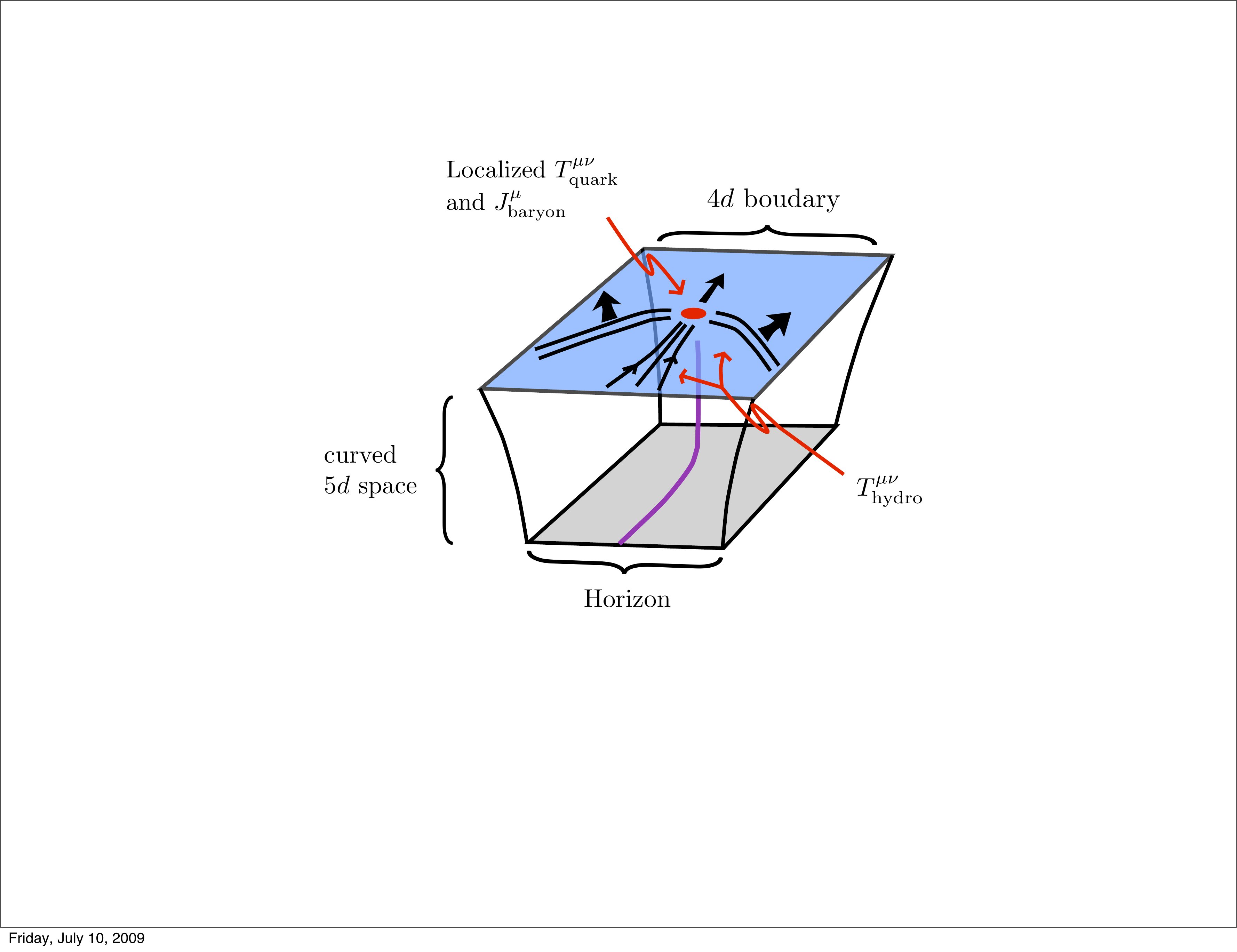}
\caption
    {\label{holographicjet}
    A cartoon illustrating the gravitational dual of a quark moving through
    a SYM plasma.  
     }
\end{figure}

Fig.~\ref{holographicjet} shows a cartoon illustrating  
the gravitational dual of a quark moving through a SYM plasma.
This consists of a string (shown as the magenta curve) moving in the AdS-BH geometry.  
The endpoints of strings are charged under a $U(1)$ gauge field 
whose influence extends to the $4d$ boundary.  The boundary itself behaves as an ideal 
electromagnetic conductor \cite{Kovtun:2005ev}, so the presence of a string endpoint
induces an \textit{image current density} on the boundary \textit{above} the endpoint.
This is illustrated schematically in Fig.~\ref{holographicjet}.
Via the standard gauge/gravity dictionary
\cite{Maldacena:1997re,Witten:1998qj} 
the induced current density corresponding to the string
endpoint has the field theory interpretation as minus the expectation value of the
baryon current density of a dressed quark.
In the limit where the string endpoint is close to the boundary the induced 
baryon density is highly localized and we may approximate 
$
\bm x_{\rm quark} \approx \bm x_{\rm string},
$
where $\bm x_{\rm string}$ is the spatial position of the string endpoint \cite{Chesler:2008wd}.

Via Einstein's equations, the presence of the string will also perturb the $5d$
geometry.  In the large $\Nc$ limit, the corresponding perturbation is order $1/\Nc^2$
and may be analyzed by linearizing Einstein's equations.  
As in the electromagnetic problem, classical field theory
implies the near-boundary perturbation in the $5d$ geometry will induce a 
corresponding perturbation in the $4d$ boundary stress tensor
\cite{Brown:1992br, Skenderis:2000in}.  Via the gauge/gravity dictionary,
the stress tensor induced on the boundary has the interpretation as the expectation value of the 
complete field theoretic stress tensor, and hence contains information about dynamics on \textit{all} length scales.

If the string endpoint is close to the boundary ---  either because of initial conditions
or the restriction of a heavy quark --- the string endpoint will have a high gravitational potential
energy.  It is this \textit{UV sensitive} part of the string energy that we identify with the energy of a quark. 
Via the gravitational equivalent of the electromagnetic image problem (\textit{i.e.} the 
gravitational bulk-to-boundary problem),
the high energy and momentum densities near the string endpoint get mapped onto
a region of $4d$ space which roughly coincides with the location of the
quark's baryon density.  That is, $T^{\mu \nu}_{\rm quark}$, as defined above in
(\ref{stressdecomp}), is \textit{induced} in the gravitational framework by the high
gravitational and kinetic energy carried by the string endpoint.  
Again, this is illustrated schematically in Fig.~\ref{holographicjet}.

With the above point in mind, a natural question arises: what is the $5d$ gravitational description
of the drag force $f^\mu$?  The answer to this question can be understood by noting that each conserved ``charge'' of the string
gets mapped into a conserved charge in the field theory.  
In the gravitational description, there in a conserved flux of energy and momentum 
which flows down the string from the endpoint.
As the energy and momentum carried by the near-endpoint portion of the string
induces $T^{\mu \nu}_{\rm quark}$, it must be that that this flux 
is related to $f^\mu$.  In the long wavelength limit,
retardation effects in the propagation of gravitational disturbances 
from the bulk to the boundary can be neglected and the locally conserved
flux of energy and momentum down the string  
--- averaged over ``microscopic'' time scales of order the horizon radius --- must in fact coincide with $f^\mu$.%
\footnote
  {
  This statement requires further explanation.  Hydrodynamics is valid on scales much greater than 
  the microscopic scale $1/T$.  Retardation effects in the bulk to boundary problem are 
  order $1/T$ and hence to leading order can be neglected in the hydrodynamic limit.  
  As noted in footnote~\ref{fave}, when one extracts $f^\mu(t)$ from hydrodynamic sources, as 
  in Eq.~(\ref{fdef}), one only extracts the drag force averaged over microscopic times.  
  Therefore, one should regard the energy-momentum flux down the string as being equivalent to $f^\mu(t)$
  only up to course graining in time.
  }

\section{Results and discussion}

\begin{figure}[h]
\centering
\includegraphics[scale=0.30]{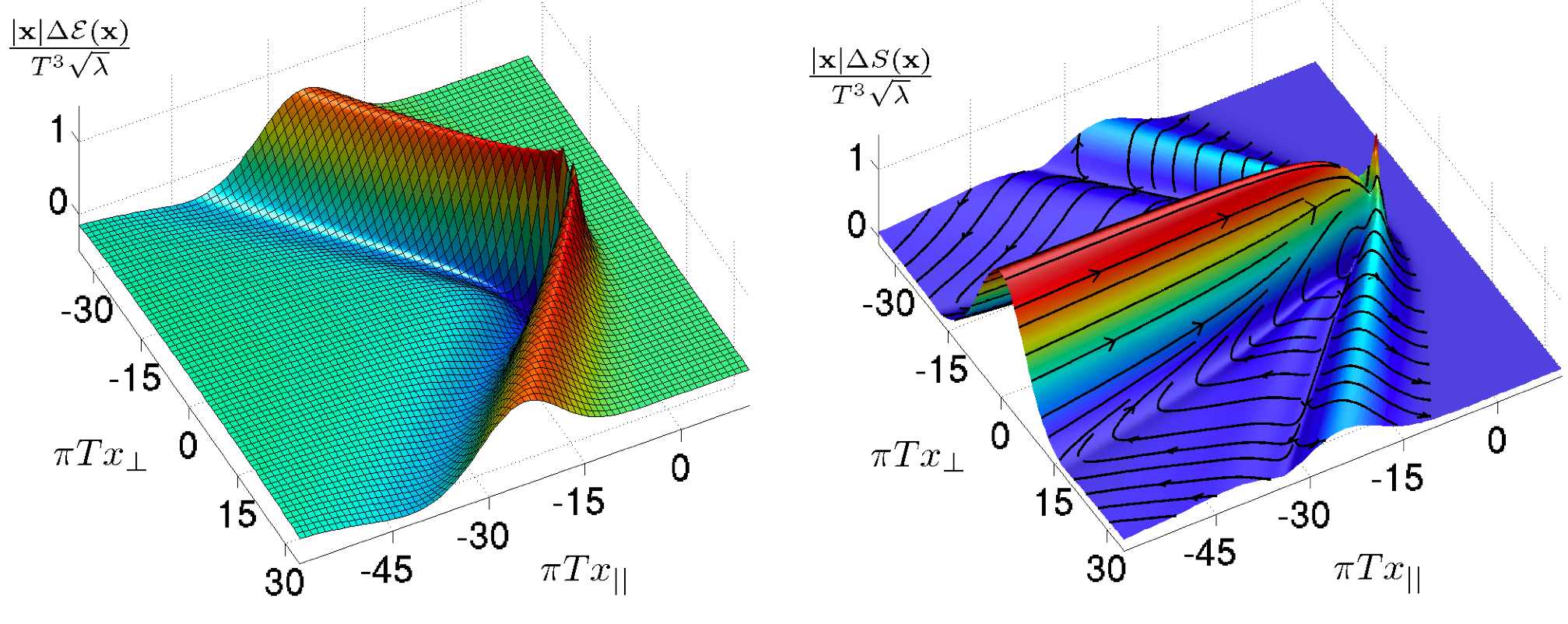}
\caption{\label{v75}
Left---Plot of energy density $| \bm x | \Delta  \mathcal E(\bm x)/(T^3 \sqrt{\lambda})$.
Right---Plot of energy flux $| \bm x | \Delta S(\bm x)/(T^3 \sqrt{\lambda})$.
In both plots the quark is moving in the $x_{||}$ direction at speed $v = 3/4$.
The flow lines on the energy flux surface are those of $\Delta\bm S( \bm x)$,
and hence indicate the direction plasma is moving.  
}
\end{figure}

Heavy quarks decelerate slowly.  Therefore, as long as one focuses on 
local questions, for sufficiently large mass it is useful to consider the limit where
the quark's velocity $v$ is constant.  Via gauge/gravity duality, the energy loss rate for a 
heavy quark moving through SYM plasma at constant velocity evaluates to 
$dE/dt = -\pi \sqrt{\lambda} T^2 v^2 /(2 \sqrt{1-v^2})$,  where $\lambda$ is the
't Hooft coupling \cite{Herzog:2006gh,CasalderreySolana:2006rq, Gubser:2006bz}.
The complete (\textit{i.e.} valid on all length scales)
energy density and energy flux of a quark moving at
constant velocity has been evaluated in 
Refs.~\cite{Chesler:2007sv, Gubser:2007ga,Gubser:2007xz,Chesler:2007an}.  
Fig.~\ref{v75}, taken from Ref.~\cite{Chesler:2007sv},
shows the complete energy density and energy flux created by a heavy quark moving at 
velocity $v = 3/4$ though a strongly coupled SYM plasma.  In a conformal theory such as 
SYM, the speed of sound is $c_s = \sqrt{1/3} \approx 0.58$, so Fig.~\ref{v75} shows supersonic motion.
As is evident in Fig.~\ref{v75}, 
a Mach cone, with opening half angle $\theta_M \approx 50^{\circ}$
is clearly visible in both the energy density and the energy flux.
Near the Mach cone, the bulk of the energy flux flow is orthogonal to the
wavefront.  Also evident in Fig.~\ref{v75} is the presence of a diffusion
wake in the energy flux.  

As emphasized in the previous section, with knowledge of the drag force
on the quark and the viscosity of SYM plasma, one can completely determine
the large-distance structure of the quark's wake.  A quantitative comparison of the complete 
quark wake to that predicted by hydrodynamics with the leading order effective source (\ref{leadingsource}) was done in Ref.~\cite{Chesler:2007sv},
and it was demonstrated that the hydrodynamic approximation provides a good description of the quark
wake even at distances $r \approx 1.5/T$ from the quark.
Therefore, essentially all of the structure seen in Fig.~\ref{v75} is simply that of hydrodynamics.  
The agreement between hydrodynamics and gauge/gravity duality
in a strongly coupled SYM plasma bolsters the notion that one should be able to model accurately the wake
produced by a high energy quark (or gluon) moving through a strongly coupled QCD plasma
merely using hydrodynamics.

\begin{figure}[h]
\centering
\includegraphics[scale=0.36]{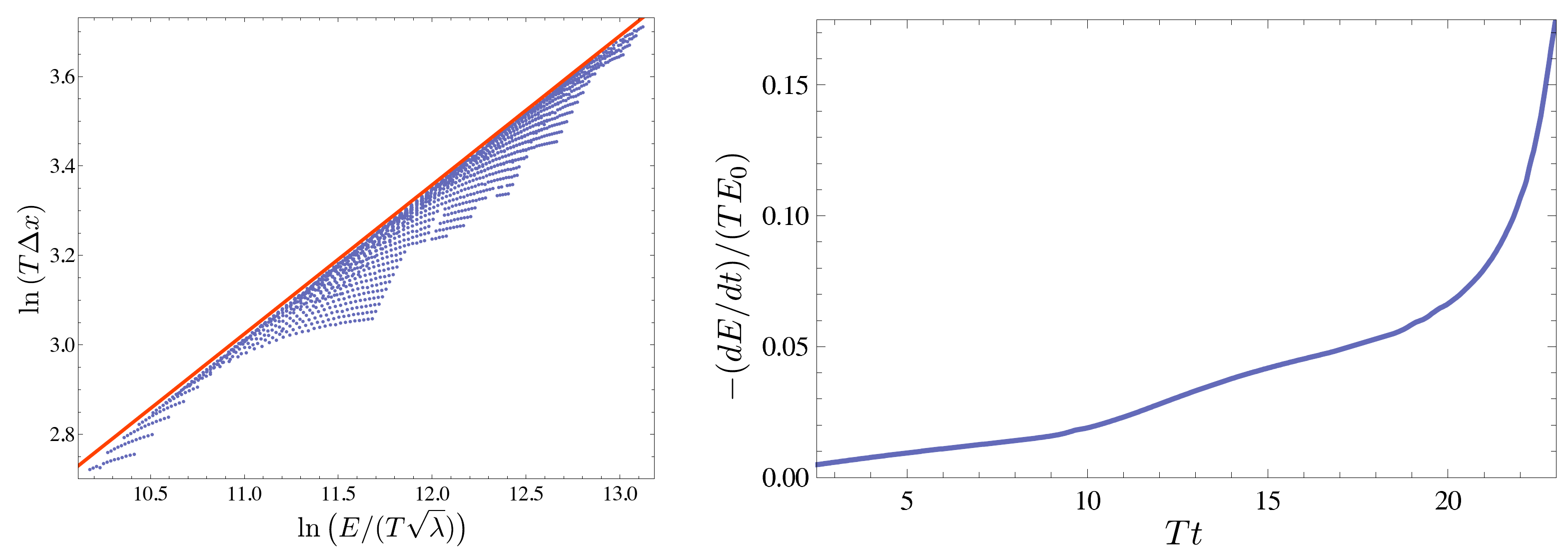}
\caption
   {
  \label{lightquark}
  Left---A log-log plot of the light quark stopping distance $\Delta x$
  as a function of total quark energy $E$
  for many falling strings with different
  initial conditions. Right---The instantaneous energy flux down a falling string  
  normalized by its initial energy $E_0$.
   }
\end{figure}

The penetration depth of a light quark (or gluon) was studied in Refs.~\cite{Chesler:2008uy,Gubser:2008as}.
In Ref.~\cite{Gubser:2008as} it was argued that the total distance a light quark
can travel scales with the cube root of its energy in the high energy limit.
Fig.~\ref{lightquark}, taken from Ref.~\cite{Chesler:2008uy},  shows the penetration depth of nearly-onshell
light quarks as a function of energy for many different sets of initial conditions and for energies $E \gg \sqrt{\lambda} T$.%
\footnote
  {
  In the gravitational setup the different sets of initial conditions for a quark simply amount to different initial conditions for a falling 
  string.  
  }
As is evident from the figure, for a given energy the total distance a quark can travel is not 
unique.  However, for a given energy there is a maximum distance a quark can travel.
This bound, shown as the solid red line in Fig.~\ref{lightquark}, is given by
$\Delta x_{\rm max} = ({0.526}/{T})\big( {E}/{T\sqrt{\lambda}} \big)^{1/3}$.

From the $E^{1/3}$ scaling of the penetration depth one might naively expect 
the light quark energy loss to scale like $d E/dt \sim E^{2/3}$.  However,
as discussed in Ref.~\cite{Chesler:2008uy} this turns out not to be the case.
To elucidate this point, also shown in Fig.~\ref{lightquark} is the instantaneous energy flux down a dual falling string
as a function of time.  The energy flux is evaluated a spatial distance $\approx 1/2 T$ from
the string endpoint.  As discussed in the previous section, the flux of four momentum
down the string is related to the rate the dual quark looses four momentum to the plasma.
More precisely, the energy flux down the string --- averaged over microscopic times, which are order $1/T$ ---
yields the four vector $f^\mu$ appearing in the effective source for hydrodynamics in Eq.~(\ref{leadingsource}).
As is evident from Fig.~\ref{lightquark}\,,the energy flux down the string does not decrease
in a power-law fashion as a naive $E^{{2}/{3}}$ scaling of $dE/dt$ would suggest,
but rather increases monotonically!

As stressed in Ref.~\cite{Chesler:2008uy}, the precise form of the energy flux down the string
is sensitive to the initial conditions used to create the string.
However, for very long-lived excitations the late time behavior of the energy flux is universal.
At late times the energy flux away from the string endpoint 
grows like $1/\sqrt{t_{\rm therm} - t}$ where $t_{\rm therm}$ is the 
thermalization time (\textit{i.e.} roughly the time where
a falling string's endpoint approaches the event horizon)  \cite{Chesler:2008uy}.%
\footnote
  {
  One should bear in mind that the thermalization time 
  is only defined with order $1/T$ accuracy.  
  }    
This late-time behavior implies that
after traveling substantial distances through the plasma,
the thermalization of light quarks
ends with a large transfer of energy to the plasma.
This behavior is qualitatively similar to the 
energy loss rate of a fast charged particle moving through 
ordinary matter, where the energy loss rate has a pronounced peak 
(known as a ``Bragg peak'') near the stopping point.
It is noteworthy to mention that this behavior is qualitatively different 
from that of heavy quarks.
At both weak and strong coupling heavy quarks prefer to deposit most of their
energy and momentum during the initial stages of their trajectory, when their 
velocity is largest \cite{Herzog:2006gh,Moore:2004tg}.


\end{document}